\documentstyle[11pt,aaspp4]{article}

\begin{document}
\title{RXTE/ASM Observations of the 35-day Cycle of Her X-1}

\author{D. M. Scott \altaffilmark{1}}
\affil{Space Science Laboratory ES-84, NASA/Marshall Space Flight
Center,Huntsville,AL 35812}

\and

\author{D. A. Leahy}
\affil{Dept. of Physics, University of Calgary, University of Calgary,
Calgary, Alberta, Canada T2N 1N4}

\altaffiltext{1}{Universities Space Research Association}
\authoremail{scott@gibson.msfc.nasa.gov, leahy@iras.ucalgary.ca} 

\begin{abstract}

Her X-1 has been monitored nearly continuously in soft X-rays (2-12 keV) 
since February 1996 by the ASM on board RXTE. We present analysis 
of these observations which include 23 contiguous 35-day cycles. 
We present the best yet average 35-day cycle lightcurve for Her X-1. 
35-day lightcurve features such as eclipses and absorption dips, and the 
alternating occurrence of the Main High and Short High states, are well 
measured. The occurrence of X-ray eclipse during the Low state is confirmed 
as is the presence of Main High states that ``turn-on'' sharply at orbital 
phases 0.23 and 0.68. The Main High state covers 35-day phase 0-0.31 and 
the Short High state covers phase 0.57-0.79, with phase 0 defined by Main 
High state turn-on. We find extended ingresses and egresses for the Short 
High state. Marching pre-eclipse absorption dips progressively become 
shallower during the Main High state. The marching pre-eclipse dip period is 
best determined from reanalysis of archival data as $1.651 \pm .002$ days if
the dip marching can be extrapolated from the Main High state into the 
Short High state or $1.646 \pm .003$ using dips within single Main High
states alone. 

\end{abstract}

%\keywords{pulsars: individual(Her X-1) --- X-rays: stars}

\noindent

\section{Introduction}

Her X-1/HZ Her is an unusual accretion-powered pulsar system exhibiting a great
wealth of phenomena. This eclipsing system contains a 1.24 second period
pulsar in a 1.7 day circular orbit with its optical companion HZ Her.
In addition, the system displays a longer 35-day cycle that was first
discovered as a repeating pattern of High and Low X-ray flux states
(\cite{gia73}). A Main High state and Short High state, lasting about
ten and five days each respectively, occur once per 35-day cycle and are
separated by ten day long Low states.
X-ray pulsations are visible during the High states but cease during the Low
states (\cite{gia73}, \cite{mihara91}). Other manifestations of the 35-day 
cycle include: pre-eclipse X-ray absorption dips that repeat at nearly the 
35-day and 1.7 day beat period (\cite{gia73}, \cite{cro80}), optical 
pulsations occurring at certain 35-day and orbital phases (\cite{mid83}), 
systematic 35-day variations in the optical orbital light curve (\cite{dee76}, 
\cite{ger76}) and High state evolution of the X-ray pulse profile 
(\cite{sco93}, \cite{dee98}). The phenomena associated with the 35-day cycle
are generally believed to arise from a tilted counter-precessing accretion 
disk that periodically obscures the X-ray emitting neutron star 
(see \cite{pried87} for a review of models of Her X-1).

Long term X-ray monitoring of Her X-1 was conducted in the early 1970's using
UHURU which observed 16 Main High states and one Short High state 
(\cite{gia73}, \cite{jon76}) 
in the 2--6 keV range. Many of the characteristic features of the 35-day 
lightcurve such as the High and Low states, the High state eclipses, 
absorption dips and the tendency of the Main High states to turn-on near 
orbital phases 0.2 and 0.7 were discovered with these observations. 
From 1974 
to 1980 Her X-1 was monitored by the Ariel-V Sky Survey Instrument for a 
total of 250 days (\cite{ric82}). Monitoring of the pulsed flux during the
Main High state has been carried out by the Burst and Transient Source
Experiment (BATSE) on board the {\it Compton Gamma Ray Observatory} (CGRO)
from April of 1993 to the present (\cite{bildsten97}). Higher resolution short 
term monitoring has been conducted with HEAO, Tenma, EXOSAT and Ginga 
(\cite{soo90}, \cite{oha84}, \cite{oge88}, \cite{sco93}, \cite{lea95})
which has confirmed many
of the X-ray phenomena first reported with the UHURU observations, added
significant details and discovered a few new phenomena such as eclipses 
during the Low state (\cite{par85}, \cite{cho97}) and an extended Low state 
in the mid-1980's (\cite{oge88}). 

Beginning in February of 1996, Her X-1 has also been monitored nearly 
continuously by the Rossi X-ray Timing Explorer (RXTE) All-Sky Monitor (ASM) 
in the 2-12 keV range.
This long term monitoring of Her X-1 has afforded a new opportunity to
study the systematics of the 35-day and orbital lightcurves. 
We examine features of the 35-day lightcurve such as the Main High
and the Short High state profiles.
The Short High state has been observed regularly now
during the 35-day cycle for the first time.
Other features are apparent,
such as the marching pre-eclipse (PE) absorption dip phenomena, 
and are described below.

\section{Monitoring by the RXTE All-Sky Monitor}

The ASM on RXTE (\cite{lev96}) 
consists of three scanning shadow cameras (SSC's) with
a net active area of $\sim 30$ $\rm cm^2$ in the 2--12 keV band per SSC. 
Each SSC has a field of view of $6^\circ$ by $90^\circ$ FWHM and is rotated
in a sequence of ``dwells'' so that the most of the sky can be covered in
one day. 
%A description of the RXTE All-Sky monitor is available at (ref).  
We examine the single dwell data from the definitive results data base.
These data consist of dwells with an exposure typically of 90 seconds 
taken once every 90 minutes on the average for a total of $\sim 20$ flux
measurements per day.  In total we include over 
13800 ASM dwells for this analysis covering the 800 day period from 
MJD 50144 to MJD 50944. 

\section{The 35-Day Cycle}

The ASM data discussed here cover 23 contiguous 35-day cycles.
The turn-on of the first cycle starts near MJD 50146.
The ASM single-dwell data were examined and show clearly the Main High
states of the 35-day cycle of Her X-1. 
The Short High states are also
detected but at much lower signal-to-noise than Main High states. 
For this paper we label the 35-day cycles consecutively as 1 to 23,
starting with the first cycle.

The ASM light curves, in many cases, were
complete enough to determine well the orbital phase of Main High state
turn-on. 
In other cases only loose or no constraints could be determined. 
We find two types of turn-on's (TO): an $\approx 0.2$ orbital phase TO 
and an $\approx 0.7$ orbital phase TO, hereafter referred to simply as
0.2 TO's or 0.7 TO's respectively.  
We also examined the BATSE light-curves of 
Her X-1 for the same 23 Main High state cycles as the ASM data. 
For cycles 1 to 11 and 17 high time resolution folded-on-board BATSE data was 
available. For the other cycles pulsed monitor 
data was used, which consists of pulsed flux and frequency measurements on a 
one-day time scale (see \cite{bildsten97} for details), thus the 
classification for these cycles is less reliable. 

A difficulty in classifying the turn-on type from the ASM observations
or other observations of low sensitivity is illustrated by cycle 17. 
From the ASM observations alone the Main High state appears to turn-on
near orbital phase 0.7. However, the higher resolution folded-on-board BATSE 
lightcurve, as well as an RXTE/PCA observation of the turn-on 
(Staubert, private 
communication) clearly show that the turn-on occured near orbital phase 0.2. 
The turn-on was followed by a large anomalous dip centered at orbital phase 
0.55. The egress from the anomalous dip is easily mistaken as the turn-on and
this mistake may have occured in classifying past turn-ons when observations 
near orbital phase 0.2 were sparse or of low sensitivity as is the case here 
with the ASM observations. For example, fig. 1 of \cite{cro80} shows a 
``clear example of a $\phi = 0.7$ turn on'', yet a small flux rise preceding
orbital phase 0.5 is apparent followed by what may be an anomalous dip. 

The combination of the ASM and BATSE lightcurves was enough to classify 
the orbital phase of TO as either 0.2 or 0.7, except for one case, 
cycle 16.
Table 1 gives the 23 cycles, a calculated turn-on time and classification 
as either an 0.2 or 0.7 turn-on, the phase jump in orbital 
cycles (i.e. the difference between the time interval from the current 
turn-on to the next one and 20.5 orbital cycles), and the cycle length. The 
turn-on time is calculated as the Modified Julian Day (MJD = JD -- 2440000.5) 
of orbital phase 0.68 for the 0.7 TO Main High states or as the time of 
orbital phase 0.23 for the 0.2 TO Main High states. 
We note that the existence of just these two types of turn-on implies
that phase jumps between successive 35-day cycles must be an integral 
multiple of 0.5 orbital cycles. 
It also implies that if 0.2 and 0.7 TO cycles alternate, on average,
and the phase jumps average to zero, 
the average 35-day cycle period must be exactly (integer plus 0.5) times
the orbital period. The values of orbital and 35-day periods give this
number as 20.5.

Over the 23 35-day cycles studied here, the net phase jump is -1 orbital
cycles, consistent with a long term average of zero.
The analysis here gives one of the best determinations so far of 
the ordering  of 0.2 TO and 0.7 TO 35-day cycles and the phase jumps.
From Table 1, 20.5 orbits is the most common 35-day cycle length
(11 cases), 21 orbits occurs 6 times, 20 orbits occurs 2 times, and
19.5 orbits occurs once (though this case is of low confidence).

We sorted the 35-day cycles into two groups according to the type of 
Main High state TO, and then averaged all of the 0.2 TO data into a single 
35-day lightcurve (Figure 1, top panel) and all of the 0.7
TO data into a second 35-day lightcurve (Fig. 1, bottom panel).
The data were phased such that the eclipses in the first orbit after
the Main High state turn-on were aligned in the averages. 
The data were binned such that the orbital phase interval of the eclipse
($\pm 0.0697$; \cite{lea95b}) occupied one bin and the remaining 
orbital phase was divided into either 7 bins (Fig. 1) or 14 bins 
(Fig. 2 and Fig. 3).

The folding and averaging of the ASM data was conducted using an orbital
period for Her X-1 of $P_{orb} = 1.7001674132$ days and an epoch
of $90^\circ$ mean longitude $T_{90} = 50086.6390977$ MJD TDB. These
values are calculated from the orbital parameters given by 
\cite{wilson97} and take into account the known orbital decay 
of Her X-1. A 35-day period given by $P_{35d} = 20.5 P_{orb}$ was used,
consistent with the value found via an epoch folding period search. 
All the times from the ASM were corrected from the geocenter to the 
solar system barycenter using the JPL ephemeris DE 200 prior to folding 
and averaging.  

The Main High state TO's are very sharp in both the 0.2 TO and 0.7 TO averaged 
light curves (see also Fig. 2), strongly supporting the idea that TO's occur 
only at 0.2 or 0.7 orbital phase with little dispersion about either value. 
From a higher time resolution average light curve we measured the turn-ons at 
orbital phases $0.23 \pm 0.02$ and $0.68 \pm 0.02$. 
The eclipses show up in the ASM light-curves clearly, which have their 
predicted ingress and egress times marked by the vertical dashed lines.

The previously known 35-day cycle features are clearly exhibited. Defining
35-day phase 0 as Main High state turn-on, the Main High state lasts from 
phase 0 to 0.31, the following Low state lasts 0.26 cycles (phase interval 
0.31-0.57),
the Short High state lasts 0.22 cycle (phase interval 0.57-0.79 following a
0.2 TO Main High state or 0.55 - 77 following an 0.7 TO Main High state),
and the second Low state lasts 0.21 cycle (phase interval $\sim 0.79-1.00$). 
The 35-day phases and average durations from the ASM data are improved 
values over previous estimates, particularly for the Short High and Low states.
The Main High and Short High state peaks are consistent with a separation of 
0.50 in 35-d phase, for both 0.2 and 0.7 TO.
The first and second Low state minima are also consistent with a 
separation of 0.50 in 35-day phase for both 0.2 and 0.7 TO's.

Fig. 2 shows the average light curves for the
0.2 TO Main High state (top panel) and the 0.7 TO Main High state (bottom panel)
at higher time resolution. 
The Main High state is seen to maintain a sharp turn-on and a gradual
turn-off for both TO types. 
The Main High state durations are the same for both TO types and both exhibit 
deep PE dips. A deep anomalous dip (AD) is consistently present 
in the first orbit of the 0.2 TO Main High state, in agreement with previous
observations (e.g. \cite{cro80}).
Individual AD's are clearly present in cycles 3, 5 and 7 and ambiguous
or indeterminate in all other 0.2 TO Main High states.
We measured the average dip center at $0.60 \pm 0.02$ orbital phase with
a duration of 0.13.  
The PE dips progress to earlier orbital phase as 35-day phase increases.
The dip ingress phase is marked by the first arrow in a pair for a dip
recurrence period of 1.651 days. The second arrow marks orbital phase 
0.813.                                                                
The main differences between the two types of Main High state are: the peak 
count rate for 0.7 TO Main High state is higher (7.5 vs. 6.5 ASM c/s); the
strong AD is present in the first orbit of the 0.2 TO Main High state.
The similarity of the features of the average light curves is otherwise 
striking. 
There is weak evidence in the 0.2 TO Main High state for two AD in the fourth 
orbit (35-day phase $\sim0.17$) at orbital phases $\sim$0.25 and $\sim$0.6, 
and for an AD in the fifth orbit at orbital phase $\sim$0.5.
For both types of Main High state, the seventh orbit shows
a sharp decrease in count rate at 35-day phase $\sim 0.29$
in declining to the Low state. 

We note that we could not determine where in 35-day phase the phase jumps
occurred, in fact the phase could drift smoothly with respect to a constant
period, but it can only be clearly measured once per cycle at Main High state 
TO. Also we could not align individual Short High states due to their
lower count rate and corresponding difficulty in determining their turn-ons.
Thus we assumed the Short High state of each 35-day cycle kept
35-day phase with the Main High state at the start of the cycle.
We also combined all of the  0.2 and 0.7 TO 35-day cycles under the assumption
that the Short High state was in phase with the following Main High state. This 
resulted in minor changes in the Short High state lightcurves.

The average light-curves for the Short High states are shown in Fig. 3 
(0.2 TO in top panel, 0.7 TO in bottom panel) at higher time resolution
than in Fig. 1.
The Short High state shows a sharp TO and slow decline similar to the 
Main High state.
The 0.7 TO Short High state is brighter than the 0.2 TO Short High
state (peak 2.8 ASM c/s vs. 2.1). 
The last orbit in the Low state just prior to the Short High state TO shows a
possible anomalous dip at orbital phase 0.5 for both 0.2 TO and 0.7 TO 
Short High state.
There are anomalous dips in most of the orbits for both type of Short High
state. The dips for both types of Short High state show no clear evidence 
for marching to earlier orbital phase as the Short High state progresses,
although this may be due to the low signal-to-noise ratio of the ASM data.
To illustrate this point, we mark the dip ingress phase by a pair of arrows.
The first arrow in a pair is for a dip recurrence period of 1.651 days. 
The second arrow marks orbital phase 0.757. 
The eclipses show extended ingresses and egresses.

\section{Orbital Cycle Light Curves}

The ASM data were folded at the orbital period of Her X-1, with 
Main High, Short High and Low state data folded separately. The
resulting light curves are shown in Figure 4.
The sharp ingress and egress of the Main High state eclipse is 
confirmed (top panel of Fig. 4). The
average effects of the marching PE dips is seen in the orbital
phase range 0.75 to 0.9.

The eclipse in the Short High state (middle panel of Fig. 4) exhibits
an extended ingress: much longer even than the ingress in the Low state.
This is confirmed in detail by RXTE/PCA observations (Scott and Leahy,
in preparation).
There is a shallow part (the part of the orbital light curve
at $\sim0.5$ ct/s) of both the ingress and egress: 
for ingress it lasts 0.32 d and for egress it lasts 0.16 d. 
AD are seen at orbital phase 0.5 during Short High state.
A previous clear example of anomalous dip during Short High state is
given by \cite{jon76}. Extended eclipse ingress appear in previous
Short High state observations as well (\cite{par80}, \cite{roc94}).
 
Low state eclipses are confirmed (bottom panel of Fig. 4). Outside of
eclipse there does not appear to be any significant modulation.
%Outside of eclipse, there is a modulation of count rate.
%It appears approximately sinusoidal with minimum of 
%0.25 c/s at phase 0.7 and maximum of 0.5 c/s at phase 0.35.
%This is likely related to changing viewing angle or obscuration of the 
%extended emission region with binary phase. 

\section{Pre-eclipse and Anomalous Dips}

Marching pre-eclipse (PE) dips are confirmed in both types of Main High state, 
as noted above. They are clearly illustrated in the ASM data in figure 2. 
The phase drift of the dips in the averaged Main High state lightcurves gives 
an approximate dip period of $1.63 \pm 0.02$ days, consistent with
either a 1.621 day beat period\footnote{The beat period is the inverse of the 
beat frequency given by $\nu_{beat} = \nu_{35d} + \nu_{orb}$} 
between the orbital and 35-day periods or the 1.65 day dip recurrence period 
measured by \cite{cro80}. We cannot determine the dip period accurately enough 
by epoch-folding individual Main High state's to discriminate between these 
two possible dip periods since the period resolution is $P^2/{2T}~=0.12$ days 
for a Main High state lasting 11 days. There is no obvious marching of the 
dips during the Short High state.

The PE dips in the ASM data exhibit the characteristic of starting out
the Main High state merged with the first eclipse ingress and progressing 
toward earlier orbital phase as the Main High state progresses. The 
PE dips have shown this same pattern since the discovery of Her X-1 
in 1971. If the PE dips were not tied to the 35-day phase then
one would expect the pattern of dips to drift significantly in orbital phase 
over many 35-day cycles. This has not been observed, so the dip pattern
appears to be tied to the 35-day ``clock''. For a dip recurrence period of 
1.65 days the fractional drift in orbital phase over one 35-day cycle is 
$\approx 0.6$ cycles, therefore requiring the dip phase to reset between 
successive Main High states. If the dip period was exactly the beat period 
then a fractional phase drift of exactly 1 orbital cycle would occur if the 
35-day period was constant, so the dip pattern would always start at the same 
orbital phase. However, as noted earlier, the time between successive 35-day 
turn-ons can vary by at least $\pm 0.5$ orbital cycles and has shown a 
longterm phase drift with respect to a constant 35-day clock (\cite{baykal93}). 

The dip pattern may be strictly tied to the time of turn-on or may be coupled 
to an underlying 35-day ``clock''. A dip pattern tied to the time of turn-on
would be expected to repeat nearly the same pattern for each 0.2 or 0.7 
turn-on Main High state. If the dip pattern is instead tied to an underlying 
35-day clock then jitter may be expected to occur in the orbital phase of the 
first dip, and hence the entire dip sequence, in a Main High state. For 
example, if the turn-on orbital phase is discretized by a thickening of the 
disk at orbital phases 0 and 0.5 then the turn-on orbital phase discretization 
and the turn-on itself would have little to due with the occurence of dips, in 
contrast to the model of \cite{cro80} where they are both considered to be
due to circulating matter at the disk rim. 
Such a disk thickening, perhaps caused by a tidal effect from HZ 
Her, would have to be tied to the position of HZ Her relative to the disk. 
The disk and dip pattern could then drift in phase with respect to a constant 
35-day clock, but the turn-on's would only show this drift in discretized 
jumps in orbital phase. Orbital phase jitter would have the effect of smearing 
out the average dip profile when averaged over many 35-day cycles. 

%The Main High state starting phase 
%of a dip pattern tied to 
%a strict beat period that is independent of the 35-day phase drift would 
%similarly show drifts in orbital phase equal to the drift in orbital
%phase between consecutive dips ($\Delta \phi_{orb} = 
%\nu_{beat} P_{orb} - 1 = 0.0488$ cycles) between consecutive Main
%High states and a longterm phase drift with respect to the eclipse if the 
%35-day cycle start phase is a random walk. Thus some phase jitter may be 
%expected to occur in the starting phase of the dip pattern and this would
%have the effect of smearing out the average dip profile when averaged over
%many 35-day cycles.  

To further investigate the PE dip period we tried an epoch folding analysis 
on the original full ASM data set, giving high period resolution. Using
the entire data set, the only significant peak present in chi-squared vs. 
folding period was at the orbital period over the period range 
1.58 to 1.75 days. The analysis was repeated using only data selected from 
either the Main High states or the Short High states. This resulted in 
significant peaks in chi-squared vs. folding period at the orbital period and 
1.621 days. Further investigation showed that the peak at 1.621 days was not
due to the presence of dips in the light curve but was due to a sampling
and aliasing effect. We created simulated data with the ASM sampling and 
errors, with orbital eclipses and a 35-day cycle envelope, but without
dips, and could reproduce the peak at 1.621 days and a folded light curve
much like that from the real ASM data. Using a sinusoidal lightcurve with
a period equal to the orbital period also resulted in peaks at 1.621 days
when only data during either the Main High state or Short High state was
selected. 

%Epoch folding the full ASM data set does not detect the dip period for
%the following reasons.
%In the absence of 35-day cycle jumps, the dip phase resets each 35-day
%cycle, in order that the dips align correctly with the first orbital cycle
%after turn-on. The phase reset is necessary since the number of dip cycles
%in a single 35-day cycle is not an integer.
%In the presence of 35-day cycle phase jumps, there is additional phase
%shift in dip phase. Either effect would make the dip recurrence 
%non-periodic on long timescales, thus remove sensitivity of epoch-folding 
%to the dip period.

The dip profile is better seen if the strong modulation of the 35-day cycle 
is removed.                                                                 
We fit a high order polynomial to the average Main High state light curve with 
dips and eclipses removed to obtain a smooth 35-day cycle envelope.
Then we construct the relative dip profile given by: 
$$ F_{rel} = (F_{actual} - F_{fitted}) / F_{fitted} + 1 $$ where
$F_{actual}$ is the actual flux and $F_{fitted}$ is the fitted flux.
This profile has a value of 0 for a 100\% reduction in 
flux from the 35-day flux envelope, and 1 for no reduction. It is shown in 
Figure 5.
The PE dips are clearly seen, with marching to earlier orbital phase.
Since we cannot measure individual dips in single 35-day cycles with the
ASM data due to the sampling, the best determination of the dip period is 
made using archival data (see Appendix)
as $1.651 \pm 0.002$ days using a phase drift method. We mark the dip ingress 
by the arrows at this period. The dips show steadily decreasing 
depth as the Main High state progresses, from 100\% reduction to $\sim50\%$ 
reduction in flux for both Main High state types.  
The depth of the PE dips is well determined except for the sixth dip in
the 0.2 TO Main High state's, since the fitted flux is a good fit except 
near the 35-day phase 0.25-0.29 where the Main High state flux changes rapidly.

The properties of the PE dips are an important factors in testing 
models for the mass transfer and accretion disk in Her X-1. In addition to the
dip recurrence period and dip depth we obtain a measure of the average dip
duration. For the anomalous dip we measure a width of $\approx 5.3$ hours.
For the PE dips we find a range in widths from 5 to 10 hours. Historical
observations of dip durations are quite sparse. Two anomalous
dip observations from \cite{cro80} give dip durations of about 2.3 and 3.4 
hours. The best case is an EXOSAT observation (\cite{voges87}) which gives a 
width of 0.7 hours. An RXTE/PCA observation of the anomalous dip in Main High 
state number 17 lasts about 6 hours (Staubert, private communication). 
Upper and lower limits on durations were determined for a large number of
PE dips by \cite{lea97}.
 
The large AD at orbital phase 0.6 in the first orbit of
many 0.2 TO Main High states is confirmed (Fig. 2).
For the Short High state, the average 35-day light curve (Fig. 3) 
shows many deep AD. The Short High state AD are consistent 
with having a period equal to the orbital period, so that the middle panel of 
Fig. 4 shows the mean AD profile.              

\section{Conclusions}

Her X-1 has been monitored nearly continuously in soft X-rays (2-12 keV) 
since February 1996 by the ASM on board XTE. We present analysis 
of these observations which include 23 contiguous 35-day cycles. 
We present the best yet determined average 35-day cycle light curve for
Her X-1. Well known 35-day lightcurve features 
such as eclipses and absorption dips, and
the alternating occurrence of the Main High and Short High states, 
are well characterized.

We determine that the Main High states 
``turn-on'' sharply at orbital phases 0.23 and 0.68, and give the sequence of 
these two types of Main High state and the phase jumps that occur during the 
ASM data. We construct mean 0.2 TO and 0.7 TO 35-day cycle light curves.
For both types of 35-day cycle, the Main High state covers 35-day phase 0-0.31 
and Short High state cover phase 0.57-0.79, with phase 0 defined by Main High 
state turn-on.
The peaks of the Main High and Short High states, and the minima of the two 
Low states both have separations consistent with 0.50 in 35-day phase.
The two main differences between the two types of 35-day cycle are: 
the peak count rate for 0.7 TO cycle is higher than the 0.2 TO
(Main High state peak of 7.5 ASM c/s vs. 6.5; Short High state peak of 2.8 ASM 
c/s vs. 2.1); AD are apparent in the 0.2 TO Main High states but not the 0.7 TO 
Main High states. The similarity of the features of the average light curves 
is otherwise striking, both in the overall flux envelope and in the timing of 
the PE dips. 

Eclipse is clearly measured in the Main High, Short High and Low states.
We find extended ingresses and egresses for the Short High state. 
PE dips are confirmed to be marching during the High State.
We refine the value of the Main High state PE dip recurrence period using 
archival data to $1.651 \pm .002$ days (or or $1.646 \pm .003$ if archival
cycle 0 is excluded). The Main High state PE dips become progressively 
shallower (from 100\% to 50\% depth) as Main High state proceeds.

\begin{acknowledgements}
DAL acknowledges support from the Natural Sciences and Engineering
Research Council of Canada. The authors are grateful for constructive
suggestions made by the referee.
\end{acknowledgements}

\appendix
\section{Appendix: Analysis of Archival Data on Pre-eclipse Dips}

We obtained archival data for dip ingress times, used by \cite{cro80},
hereafter referred to as CB80, and supplemented it with some Short High state 
dip measurements by \cite{ric82} using Ariel V and VI data. 
This data is plotted in a dip phase vs. 35-day phase diagram in Fig. 6.
When dip phase is defined to be the 1.7 day orbital period (top panel)
the data points lie on a pair of downward sloped lines 
(one for PE dips, one for anomalous dips). 
For dip phase defined as 1.621 day beat period (bottom panel), 
the lines are  upward sloped.
For both cases the phase drift is consistent with the 1.65 day period
found below and by CB80. % \cite{cro80}.
However when dip phase is defined as 1.65 day period, a scatter diagram results
(Fig. A1, middle panel).
This is explained as follows: the period within a single 35-day 
cycle is 1.65 day, but phase jumps between 35-day cycles 
cause what would otherwise be a single horizontal set of points in Fig. 6
to be split into a set of horizontal sets (one for each 35-day cycle) offset 
by the phase jumps, i.e. indistinguishable from a scatter diagram. We
illustrate this by denoting dip ingress times from a single Main High state
with asterisks in the middle panel of Fig. 6. 

To determine the period in the presence of phase jumps we need to
examine data within each single Main High state separately.
There are 6-7 PE dips detectable during each Main High state. The dip 
recurrence period within each individual Main High state can be determined by 
epoch folding, however the period resolution is $P^2/{2T}~=0.12$ days,
larger than the difference between the 1.65 day and 1.621 day periods. 

Thus we employ a phase drift method, which is more sensitive. 
The drift between 1.621 days and 1.65 day periods over 6 dips is $0.174$ days, 
or $\approx 0.1$ dip cycles. The dip ingress times of 11 Main High state's 
were examined individually. This avoids the problem with the CB80 analysis of 
mixing all the data from many Main High state's together. 
The dip recurrence period was determined by fitting a line to the dip 
ingress times in a time vs. orbital phase diagram (equivalent to 35-day phase 
vs. orbital phase when using data from a single 35-day period). 
We define: $\Delta \phi$ is the change in orbital phase between successive 
dips $(<0)$, and $\Delta t$ is the change in time between successive dips 
$(>0)$. Then:

$$ \frac{\Delta \phi}{\Delta t} = (\frac{P_{dip} -
P_{orbit}}{P_{orbit}})(\frac{1}{P_{dip}})  $$

$$ P_{dip} = \frac{P_{orb}}{1 - P_{orb}(\frac{\Delta \phi}{\Delta t})} $$

$$ \sigma_{dip} = \sigma_{(\frac{\Delta \phi}{\Delta t})} \vert \frac{
P_{orb}^2}{(1 - P_{orb} (\frac{\Delta \phi}{\Delta t}))^2} \vert  $$

$ \frac{\Delta \phi}{\Delta t}$ can be determined from slope of linear fit
to dip ingress times in the time versus orbital phase diagram.
The resulting individual determinations of dip period from the 11 35-day cycles
are plotted  in Figure 7.
For the 11 Main High state's we get an average dip period of 
$1.651 \pm 0.002$ days. The dip ingress recurrence period is thus different
from the 1.621 day beat period with a significance of 14 sigma. 
The dip recurrence period determined by CB80 was $1.65 \pm 0.006$ days 
so we have improved the error by a factor of three. The dip recurrence
period from cycle 0 has the smallest error due to the assumption, made by
CB80, that the dip marching can be extrapolated from the Main High state into 
the following Short High state. All other points use only data within
a single Main High state. If we assume this extrapolation is unwarranted and 
delete the point from cycle 0, then we find an average dip recurrence period of
$1.6459 \pm 0.0034$ days. The dip recurrence period is consistent with 
1.65 days in either case and to date no data set of a quality and coverage 
comparable to the UHURU data is available to check this result.
We also examined the phases of the dips between different 35-day cycles
from the intercepts of the best-fit line to the dips in the 35-day
phase vs. orbital phase diagram. We find the line intersects 35-day phase
0 in a narrow range of orbital phase: a mean value of 0.859 and variance
0.022.

%avg dip period        1.6458828 +/-     0.0034109100  w/o cycle 0
%avg dip period        1.6506779 +/-     0.0020730666  w/  cycle 0

\newpage

\begin{deluxetable}{ccccc}   % {lrrr} % rcrrrrr}
\small
\tablenum{1}
\tablewidth{0pt}
\tablecaption{35-day Cycle Turn-on Times, Types, Phase Jumps and Lengths}
\tablehead{
%\colhead{Cycle \#}             & \colhead{Turn-on Type}          &
%\colhead{Phase Jump (orbits)}  & \colhead{Cycle Length (orbits)} 
\colhead{Cycle}      & \colhead{Turn-on time} & \colhead{Turn-on}          &
\colhead{Phase Jump} & \colhead{Cycle Length}   \nl
\colhead{\#}        & \colhead{MJD}  &\colhead{Type}          &
\colhead{(orbits)}  & \colhead{(orbits)}  
}
%\cutinhead{\# & Type & (orbits) & (orbits)}
%\tablecolumns{4} 
\startdata
 1 & 50146.54  & 0.2 & 0    & 20.5 \nl
 2 & 50181.30  & 0.7 & 0    & 20.5 \nl
 3 & 50216.24  & 0.2 & 0.5  & 21   \nl
 4 & 50251.95  & 0.2 & -0.5 & 20   \nl
 5 & 50285.95  & 0.2 & 0    & 20.5 \nl
 6 & 50320.72  & 0.7 & 0    & 20.5 \nl
 7 & 50355.66  & 0.2 & 0    & 20.5 \nl
 8 & 50390.42  & 0.7 & 0    & 20.5 \nl
 9 & 50425.36  & 0.2 & 0    & 20.5 \nl
10 & 50460.13  & 0.7 & 0.5  & 21   \nl
11 & 50495.84  & 0.7 & 0    & 20.5 \nl
12 & 50530.77  & 0.2 & 0    & 20.5 \nl
13 & 50565.54  & 0.7 & 0.5  & 21   \nl
14 & 50601.25  & 0.7 & 0.5  & 21   \nl
15 & 50636.95  & 0.7 & ?    & ?    \nl
16 & ?         & ?   & ?    & ?    \nl
17 & 50704.19  & 0.2 & 0.5  & 21   \nl
18 & 50739.89  & 0.2 & -0.5 & 20   \nl
19 & 50773.90  & 0.2 & 0.5  & 21   \nl
20 & 50809.60  & 0.2 & -1.0 & 19.5 \nl
21 & 50842.67  & 0.7 & 0    & 20.5 \nl
22 & 50877.61  & 0.2 & 0    & 20.5 \nl
23 & 50912.38  & 0.7 & ?    & ?    \nl
\enddata
\end{deluxetable}

\newpage
 
Fig. 1. The XTE ASM 2--12 keV average 35-day light curve of Her X-1.
Vertical dashed lines mark eclipse ingress and egress. 35-day phase
0 is set at orbital phase 0.25 for the 0.2 Turn-on lightcurve
and at orbital phase 0.75 for the 0.7 Turn-on lightcurve.
Top panel: 0.2 Main High state turn-ons.
Bottom panel: 0.7 Main High state turn-ons.

Fig. 2. The XTE ASM 2--12 keV average light curve of the Main High 
state. Vertical dashed lines mark eclipse ingress and egress. 
Arrow pairs mark the phase of predicted dip ingress according to a dip
recurrence period of either 1.651 days or the 1.7 day orbital period.     
Top panel: 0.2 Main High state turn-ons.
Bottom panel: 0.7 Main High state turn-ons.

Fig. 3. The XTE ASM 2--12 keV average light curve of the Short High
state. Vertical dashed lines mark eclipse ingress and egress. 
Arrow pairs mark the phase of predicted dip ingress according to a dip
recurrence period of either 1.651 days or the 1.7 day orbital period.     
Top panel: 0.2 Main High state turn-ons.
Bottom panel: 0.7 Main High state turn-ons.

Fig. 4. Orbital lightcurves of Her X-1 from folding the XTE ASM 2--12
keV data. The topmost panel shows the Main High state, the middle panel 
shows the Short High state and the bottom panel shows the Low state. The
vertical dashed lines marked orbital phases 0.9303 and 0.0697 corresponding to
eclipse ingress and egress during the Main High state.
   
Fig. 5. Relative dip profile, $F_{rel}$, during the Main High state showing 
eclipses and marching PE dips. Vertical dashed lines mark eclipse ingress and 
egress. Arrows mark the phase of predicted dip ingress 
according to a dip recurrence period of 1.651 days. 
Top panel: 0.2 TO Main High state. 
Bottom panel: 0.7 TO Main High state.

Fig. 6. Archival dip ingress times plotted in a dip phase vs. 35-day phase 
diagram. Top panel for dip period equal to orbital period; Middle panel for 
1.651 day dip period; Bottom panel for 1.621 day dip period. Asterisks in 
middle panel denote dip times from {\it archival} 35-day cycle \#3. 

Fig. 7. Dip recurrence periods determined by the phase drift method,
for 11 35-day cycles of archival dip ingress times obtained from archival
observations. The first cycle combines Main High and Short High state dip 
times, all others use Main High state dips only. The average period and its 
error are shown by the horizontal solid and dashed lines near 1.651 day. 
Cycle 0 starts in December 1971.
 
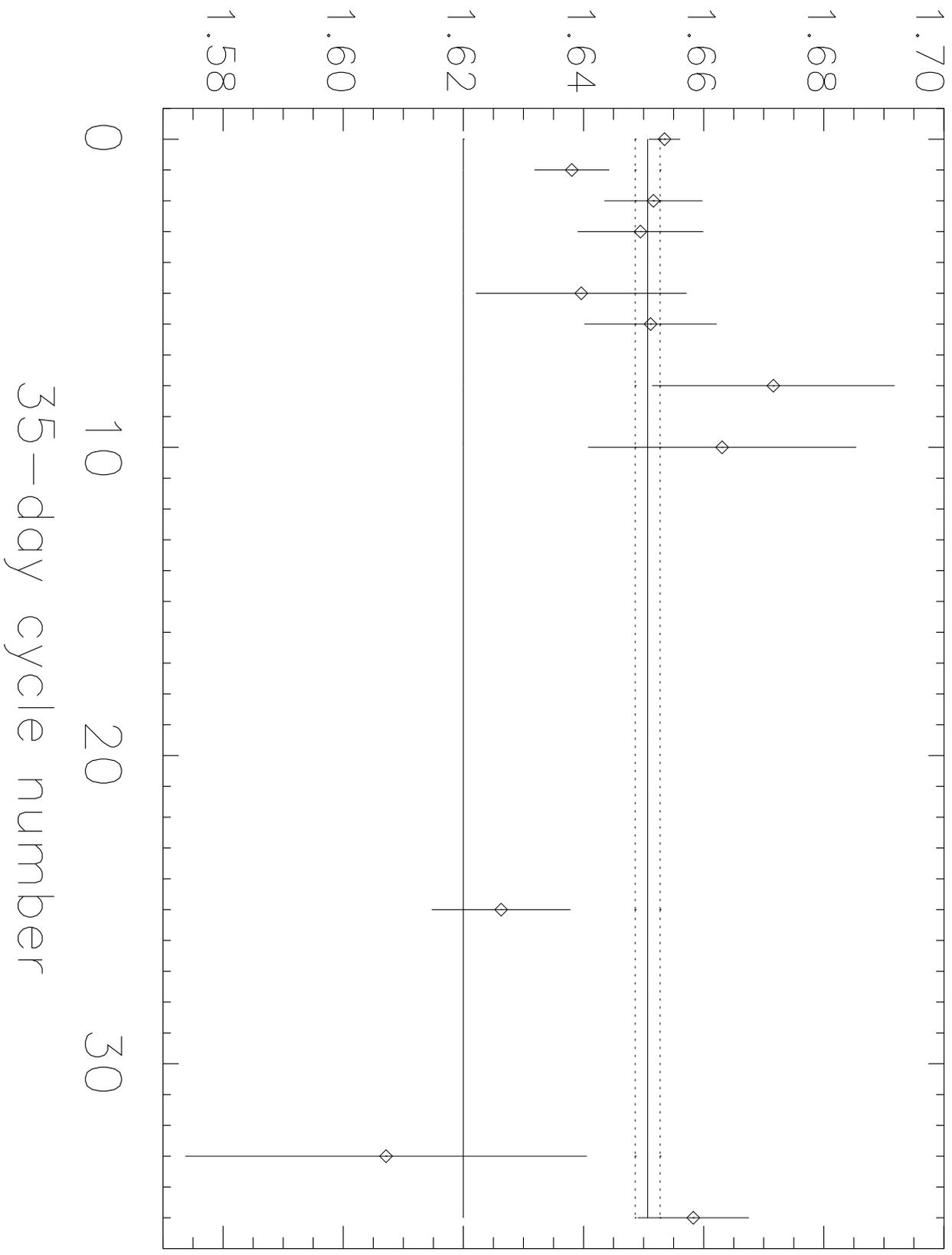
\end{document}